\documentclass[referee] {aa}
\usepackage{graphics,epsfig,here,txfonts}

\begin{document}
\title{Search for low energy $\nu$ in
correlation with the 8 events observed by the EXPLORER and
NAUTILUS detectors in 2001}
\titlerunning{LVD and GW antennas}
\authorrunning{LVD Collaboration}
\author{
M.Aglietta\inst{1}, P.Antonioli\inst{2}, G.Bari\inst{2},
C.Castagnoli\inst{1}, W.Fulgione\inst{1}, P.Galeotti\inst{1},
M.Garbini\inst{2}, P.L.Ghia\inst{1}, P.Giusti\inst{2},
F.Gomez\inst{1}, E.Kemp\inst{3}, A.S.Malguin\inst{4},
H.Menghetti\inst{2}, A.Porta\inst{1}, A.Pesci\inst{2},
I.A.Pless\inst{5}, V.G.Ryasny\inst{4}, O.G.Ryazhskaya\inst{4},
O.Saavedra\inst{1}, G.Sartorelli\inst{2}, M.Selvi\inst{2},
D.Telloni\inst{1}, C.Vigorito\inst{1}, L.Votano\inst{6},
V.F.Yakushev\inst{4}, G.T.Zatsepin\inst{4}, A.Zichichi\inst{2}}
\offprints{W.Fulgione Istituto di Fisica dello Spazio
Interplanetario, corso Fiume 4, 10133-Torino (I)}
\mail{FULGIONE@TO.INFN.IT} \institute{ IFSI-CNR,Torino, University
of Torino and INFN-Torino, Italy \and University of Bologna and
INFN-Bologna, Italy \and University of Campinas, Campinas, Brazil
\and Institute for Nuclear Research, Russian Academy of Sciences,
Moscow, Russia \and Massachusetts Institute of Technology,
Cambridge, USA \and INFN-LNF, Frascati, Italy }
\date{Received xx-yy-zz, Accepted xx-yy-zz}
\abstract{
We report on a search for low-energy neutrino
(antineutrino) bursts in correlation with the 8 time coincident events observed by the gravitational
waves detectors EXPLORER and NAUTILUS (GWD) during the year 2001.

The search, conducted with the LVD detector (INFN Gran Sasso
National Laboratory, Italy), has considered several neutrino
reactions, corresponding to different neutrino species, and a wide
range of time intervals around the (GWD) observed events. No
evidence for statistically significant correlated signals in LVD
has been found.

Assuming two different origins for neutrino emission, the cooling
of a neutron star from a core-collapse supernova or from
coalescing neutron stars and the accretion of shocked matter, and
taking into account neutrino oscillations, we derive limits to the
total energy emitted in neutrinos and to the amount of accreting
mass, respectively.  }
\maketitle

\section{Introduction}
The analysis of the data collected in coincidence by the
gravitational wave bar detectors EXPLORER and NAUTILUS during the
year 2001 (Astone et al., 2002) shows an excess (8 events against
2.6 expected from the background) when the two detectors are
favorably oriented with respect to the Galactic Disc. Moreover,
this result comes from the present day most sensitive experiments
for the detection of gravitational wave bursts and a search for
neutrino bursts in correlation with the 8 GWD events is,
therefore, appropriate.

A few astrophysical transient sources are indeed expected to
produce associated bursts of neutrinos and gravitational waves. It
is well known that most of the energy ($99\%$) released in the
gravitational core collapse of a massive star is carried away by
neutrino originated both from the matter accretion in the shock
and from the cooling of the proto-neutron star (see for example
Burrows et al., 1992). Depending on the collapse dynamics, some
fraction of the total energy is emitted in GW ( Thorne 1988,
Muller 1997 ), asymmetric supernovae in our Galaxy being the best
candidate sources for GW bar detectors. Two coalescing neutron
stars would  also constitute a source for both neutrinos and
gravitational waves. From the point of view of GW emission, it is
likely that the merging event would produce powerful gravitational
wave bursts, and, even if the physics of the merger is not known,
there are estimates that, for binary systems of large mass,
coalescence waves are
likely to be stronger than the inspiral ones.
Some amount of the kinetic energy is converted in thermal energy so that the hot remnant would probably emits thermal neutrinos.\\
The search for a neutrino burst associated to the events detected
by the GWD EXPLORER and NAUTILUS in 2001 has been performed with
the LVD apparatus, operating in the INFN Gran Sasso National
Laboratory (Italy) since 1992 with the main purpose of searching
for neutrinos from gravitational stellar collapses within the
whole Galaxy.

The paper is planned as follows: in Sect.2 we briefly describe the
LVD detector, and we explain the selection of the data. In Sect.3
we present the results of the analysis: a time interval spanning
from 12 $h$ preceding each of the 8 events up to 12 hours later,
has been scanned, searching for any excess over the statistical
fluctuation of the background. Further, a search for a $\nu$
signal coincident in time with every event has been performed. We
conclude in Sect.4, where we discuss the results of the search,
taking into account $\nu$ oscillations, and considering the two
processes for $\nu$ emission, i.e., cooling and accretion. Since
we do not find any neutrino burst candidate associated with the 8
GWD mentioned events, in the following assumed scenarios we derive
upper limits:
\begin{itemize}
\item on the neutrino flux, without reference to any particular source,
\item  on the total amount
of energy emitted in neutrinos, in the cooling case,
\item on the accreting mass, in the accretion case.
\end{itemize}

\section{The LVD Experiment and the Data Selection}
The Large Volume Detector (LVD) in the INFN Gran Sasso National
Laboratory, Italy, consists of an array of 840 scintillator
counters, 1.5  m$^3$ each, interleaved by streamer tubes, arranged
in a compact and modular geometry (see Aglietta et al., 1992, for
a detailed description). The active scintillator mass is $M=1000$
t. The experiment has been taking data, under different larger
configurations, since 1992, and reached its final
one during 2001.\\
There are two subsets of counters: the external ones ($43 \%$),
operated at energy threshold ${\cal E}_h\simeq 7$ MeV, and inner
ones ($57 \%$), better shielded from rock radioactivity and
operated at ${\cal E}_h\simeq 4$ MeV. To tag the delayed $\gamma$
pulse due to $n$-capture, all counters are equipped with an
additional discrimination channel, set at a lower threshold,
${\cal E}_l\simeq 1$ MeV.\\
The main purpose of the telescope is the detection of neutrinos
from gravitational stellar collapses in the Galaxy. In the
following we will focus on $\nu$ reactions with free protons and
$^{12}\mathrm{C}$ nuclei, constituting the bulk of the expected
signal and having the best signature in the detector, namely:
\begin{itemize}
\item \noindent {\em (1)}\ inverse $\beta$-decay: $\bar\nu_e p,e^+
n$, observed through a prompt signal from ${e}^+$ above threshold
${\cal E}_h$ (detectable energy $E_d \simeq E_{\bar\nu_{e}}-1.8$
MeV $+ 2 m_e c^2 $), followed by the signal from the ${n p,d}
\gamma$ capture ($E_{\gamma} = 2.2$ MeV), above ${\cal E}_l$ and
with a mean delay $\Delta t \simeq 180~\mu \mathrm{s}$.
\item
\noindent {\em (2)}\ $\nu_i$ and $\bar\nu_i$ neutral current
interactions with $^{12}\mathrm{C}$: $\stackrel{\scriptscriptstyle
(-)}{\nu}_i$ ${}^{12}{\rm C},\stackrel{\scriptscriptstyle
(-)}{\nu}_i$ ${}^{12}{\rm C}^*$ ($i=e,\mu,\tau$), whose signature
is the monochromatic photon from carbon de-excitation
($E_{\gamma}=15.1$ MeV), above ${\cal E}_h$.
\item \noindent {\em
(3$'$)}\ $\nu_e$ charged current interactions with
$^{12}\mathrm{C}$: $\nu_e\,^{12}\mathrm{C},^{12}\!\mathrm{N}\
e^-$, observed through two signals: the prompt one due to the
$e^-$ above ${\cal E}_h$ (detectable energy $E_d \simeq
E_{\nu_e}- 17.3 $  MeV) followed by the signal, above ${\cal E}_h$,
from the $\beta^+$ decay of $^{12}\!\mathrm{N}$ (mean life time
$\tau = 15.9$ ms). \item \noindent {\em (3$''$)}\ $\bar\nu_e$
charged current interactions with $^{12}\mathrm{C}$: $\bar\nu_e\,
^{12}\rm{C},^{12}\!\rm{B}\ e^+$, observed through two signals: the
prompt one due to the $e^+$ (detectable energy $E_d \simeq
E_{\bar\nu_e}- ~ 14.4    \, \mathrm{MeV} + 2m_e c^2$) followed by the
signal from the $\beta^-$ decay of $^{12}\!{\rm B}$ (mean life
time $\tau= 29.4$ ms). As for reaction {\em (3$'$)}, the second
signal is detected above the threshold ${\cal E}_h$.
\end{itemize}
After being subjected to a preliminary process to reject muons,
the raw data are grouped in three different classes, with specific
signatures to tag the different described reactions:
\begin{enumerate}
\item IBD class (inverse beta decay): pulses with $E_\mathrm{d}
\geq {\cal E}_h$ followed by a delayed ($\Delta t \leq 1
\mathrm{ms}$) low energy ($E_\mathrm{d}>{\cal E}_l$) pulse in the
same counter. The efficiency in tagging the n-capture is
$\epsilon=60$\% for the core counters, $\epsilon=50$\% for the
whole detector;
\item NC class (neutral current): pulses with $11$
 $\mathrm{MeV} \leq E_\mathrm{d} \leq 17.5$ MeV, the efficiency in
tagging the $\gamma$ from $^{12}\mathrm{C}$ de-excitation being
$\epsilon=55$\%
\item CC class (charged current): two pulses with
$E_\mathrm{d} \geq {\cal E}_h$ MeV, within $\Delta t \leq 150
\mathrm{ms}$, in the same counter. The efficiency in tagging the
$^{12}\mathrm{N}$ and $^{12}\mathrm{B}$ decay are $\epsilon=85$\%
and $\epsilon=70$\%, respectively.
\end{enumerate}
\section{The Analysis}
The LVD detector is sensitive to neutrino bursts from core
collapse supernovae within the whole Galaxy (Aglietta et al.,
1992). The scintillator counting rate is continuously monitored:
all the events are examined on-line on the basis of their time
sequence. Neutrino burst candidates are identified as clusters of
scintillator counter pulses with an imitation frequency less than
a predefined threshold (Fulgione, Mengotti and Panaro, 1996).
During the year 2001, no neutrino burst candidate has been
evidenced, thus allowing to conclude that no $\nu$ signal from
gravitational stellar collapse in the Galaxy has
been detected (Aglietta et al., 2003).\\
However, the absence of candidates in the LVD detector taken alone
does not preclude the possibility of positive effects, when
combining it with another detector, since the joint measurement
allows to increase the sensitivity. The analysis in correlation
with the 8 candidate events has then been conducted, in four steps
described in the following.
\subsection{Step 1. Check of the detector stability.}
First of all, the LVD detector performance at the occurrence
of the 8 GW events (see the list in Astone et al., 2002) has been
checked by studying the behavior of the counting rate in a 24
hours interval around the time of each of them.
\begin{figure}[H]
\vspace{-1.8cm}
\begin{center}
\mbox{\epsfig{file=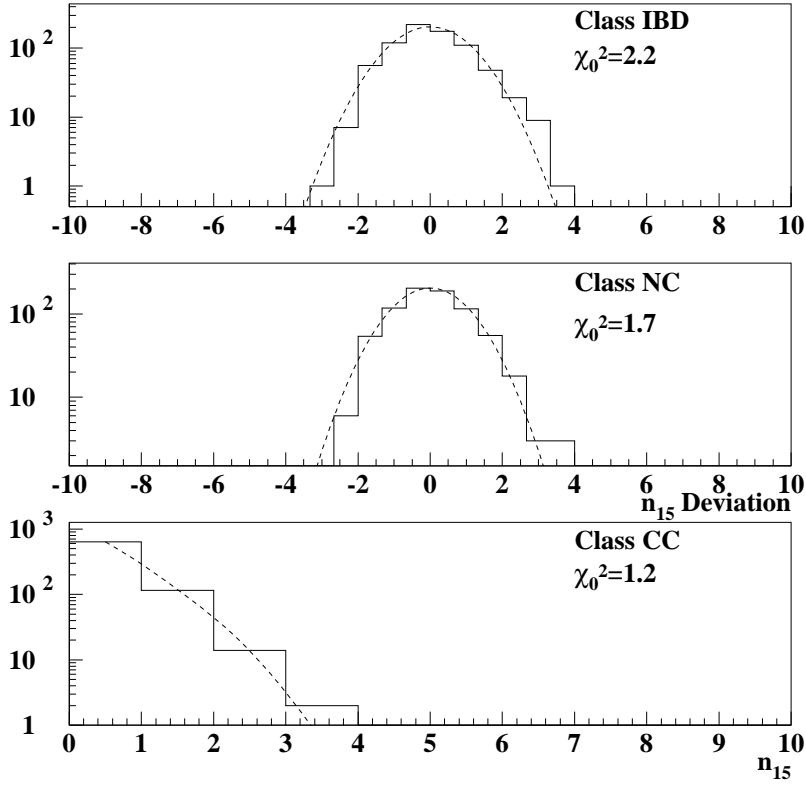,height=12cm,width=12cm}}
\vspace{-0.8cm} \caption{Distributions of the deviations of
$n_{15}$ (see text), with respect to $\langle n_{15}\rangle$, for
events of IBD class (top) and NC class (middle), and $n_{15}$
distribution, for CC class (bottom). The superimposed dotted lines
are the fits to zero mean and unit-width gaussians (IBD and NC)
and poissonian (CC).} \label{distrsd}
\end{center}
\end{figure}
For each event, the 24 hr average of the number of counts every 15
minutes, $\langle n_{15}\rangle$, is evaluated: including all the
8 events, we show in figure \ref{distrsd} the distributions of the
deviations of $n_{15}$, with respect to $\langle n_{15}\rangle$,
for events of IBD class (top) and NC class (middle); for CC class
(bottom), due to the small $<n_{15}>$, we directly show the
$n_{15}$ distribution. They have been fitted with zero mean and
unit-width gaussians (IBD and NC) and poissonian (CC): they are
shown superimposed in the same figures, together with the
resulting reduced $\chi^2$ values. The LVD counting rate, for all
the 8 events and all the data classes, is then well understood in
terms of poissonian statistics: this sets a firm base for the
following steps.
\subsection{Step 2. Search in a sliding window.}
The search for a possible $\nu$ burst has been conducted in a 24
hours interval $T$  around the occurrence of each of the 8 events.
The 8 intervals have been scanned through a ``sliding window'' of
variable duration: more in detail, they have been divided into
$N_{\delta t} = 2 \cdot \frac{T}{\delta t }-1$  intervals of
different duration $\delta t$, each one starting at the middle of
the previous one. The multiplicity distributions of clusters
(i.e., the number of events within each $\delta t$) have then been
studied for the three classes of data and for $\delta t
=1,5,10,20,50,100$ s, and have been compared with the expectations
from poissonian fluctuations of the background.
\begin{figure}[H]
\vspace{-0.8cm}
\begin{center}
\mbox{\epsfig{file=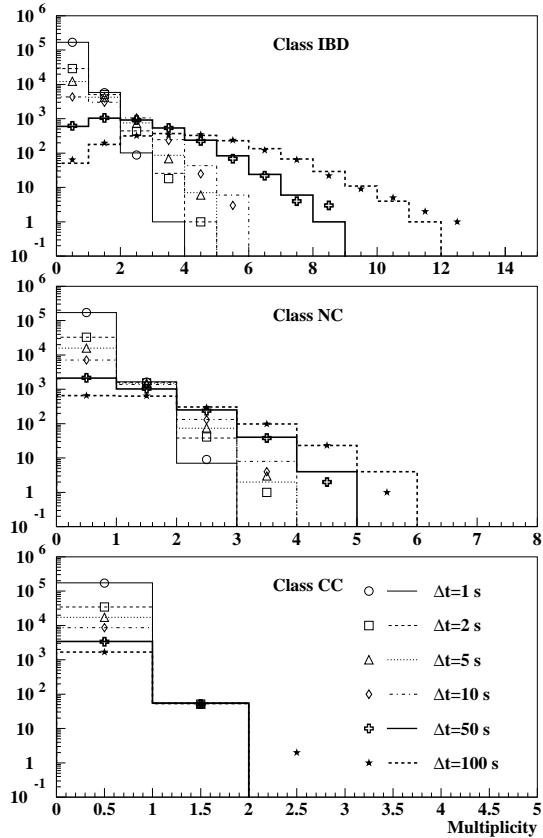,height=12cm,width=8cm}}   
\vspace{-0.4cm} \caption{Distributions of cluster multiplicities,
for IBD class events (top), NC class (medium), CC class (bottom),
together with expectations from poissonian fluctuations, in the
case of GW event n.5.} \label{distrmult}
\end{center}
\end{figure}
We show as an example the case of GW event n.5: the distributions
of cluster multiplicity, for events of IBD class (top), NC class
(middle) and CC class (bottom), in the case of the 6 different
$\delta t$, can be seen in fig. \ref{distrmult}, together with the
expectations from poissonian fluctuations of the background, the
relative reduced $\chi^2$ values ranging from 0.1 to 1.2. The
agreement between data and expectations holds also in the case of
the other seven events. This, together with the check of the
poissonian probabilities associated to each measured multiplicity,
in each class and for each event, allows to state that there is no
evidence for any detectable $\nu$ signal in correspondence of any
of the considered events.
\subsection{Step 3. Search in a fixed window.}
The search for a $\nu$ signal in coincidence with every GW event
has been further conducted using a ``fixed window'' centered at
the time of each of them. In particular, for each data class, we
compare the number of pulses ($N_\mathrm{d}$), recorded during
time windows of different duration $\delta t$, centered on each
event time, with the average number of pulses expected from
background, $N_{\mathrm{bk}}$. The value of $N_{\mathrm{bk}}$ is
evaluated by using the rate in the 24 hours around each event,
excluding the contribution of the central portion of time to avoid
the contamination due to a possible signal. Results corresponding
to $\delta t = 1,5,10,20,50,100~s$, for each of the 8 events, are
summarized in table \ref{tb1}, \ref{tb2} for IBD and NC classes,
respectively. The table relative to CC class is not shown since in
all the cases $N_\mathrm{d}=0$, and $N_{\mathrm{bk}}<3\cdot
10^{-2}$.
\begin{table}[H]
\vspace{-0.1cm} \caption{IBD class: number of events detected in
coincidence with the 8 GW events, for different durations of the
time window ($\delta t$), compared with the expectations from the
background. The effective LVD mass, $M$, at the time of each event
is also shown.}
\begin{center}
\vspace{-0.2cm}
\begin{tabular}{| c || c c c c c c c |}
\hline Ev. n. & & $ \delta t = 1$ s & $\delta t = 5$ s & $\delta t
= 10$ s & $\delta t = 20$ s &$\delta t = 50$ s&$\delta t
= 100$ s \\[-0.10cm]
($M$) & & & & & & & \\\hline
 1 & $N_\mathrm{d}$& 0&  1&  1&  1&  4&  6\\[-0.25cm]
(740t) &$N_{\mathrm{bk}}$&  $5\cdot 10^{-2}$ &  0.2&  0.5&  1.0&  2.4&  4.8\\
 \hline
 2 & $N_\mathrm{d}$& 0&  0&  0&  1&  4&  7\\[-0.25cm]
(740t) &$N_{\mathrm{bk}}$&  $5\cdot 10^{-2}$ &  0.2&  0.5&  1.0&  2.5&  4.9\\
 \hline
 3 & $N_\mathrm{d}$& 0&  1&  1&  1&  2&  6\\[-0.25cm]
(728t) &$N_{\mathrm{bk}}$&  $4\cdot 10^{-2}$ &  0.2&  0.4&  0.9&  2.1&  4.3\\
 \hline
 4 & $N_\mathrm{d}$& 0&  0&  0&  0&  1&  2\\[-0.25cm]
(726t) &$N_{\mathrm{bk}}$&  $5\cdot 10^{-2}$ &  0.2&  0.5&  0.9&  2.3&  4.6\\
 \hline
 5 & $N_\mathrm{d}$& 0&  0&  1&  3&  4& 11\\[-0.25cm]
(666t) &$N_{\mathrm{bk}}$&  $3\cdot 10^{-2}$&  0.2&  0.3&  0.7&  1.7&  3.4\\
 \hline
 6 & $N_\mathrm{d}$& 0&  0&  0&  0&  1&  2\\[-0.25cm]
(786t) &$N_{\mathrm{bk}}$&  $4\cdot 10^{-2}$&  0.2&  0.4&  0.9&  2.2&  4.4\\
 \hline
 7 & $N_\mathrm{d}$& 0&  1&  1&  2&  2&  2\\[-0.25cm]
(364t) &$N_{\mathrm{bk}}$&  $2\cdot 10^{-2}$&  0.1&  0.2&  0.3&  0.8&  1.6\\
 \hline
 8 & $N_\mathrm{d}$& 0&  0&  0&  0&  0&  0\\[-0.25cm]
(364t) &$N_{\mathrm{bk}}$&  $2\cdot 10^{-2}$&  0.1&  0.2&  0.3&  0.8&  1.6\\
 \hline
 \hline
 $\Sigma_1^8$ & $N_\mathrm{d}$& 0&  3&  4&  8&  18&  36\\[-0.25cm]
$\bar M$=639t &$N_{\mathrm{bk}}$&  0.3&  1.5& 3.0& 5.9& 14.9&  29.7\\
\hline
\end{tabular}
\label{tb1}
\end{center}
\end{table}
\begin{table}[H]
\vspace{-0.1cm} \caption{NC class: number of events detected in
coincidence with the 8 GW events, for different durations of the
time window ($\delta t$), compared with the expectations from the
background. The effective LVD mass, $M$, at the time of each event
is also shown.}
\begin{center}
\vspace{-0.2cm}
\begin{tabular}{| c || c c c c c c c |}
\hline Ev. n. & & $ \delta t = 1$ s & $\delta t = 5$ s & $\delta t
= 10$ s & $\delta t = 20$ s &$\delta t = 50$ s&$\delta t
= 100$ s \\[-0.10cm]
($M$) & & & & & & & \\\hline
 1 & $N_\mathrm{d}$& 0&  0&  0&  0&  1&  3\\[-0.25cm]
(740t) &$N_{\mathrm{bk}}$&  $1\cdot 10^{-2}$ &  $6\cdot 10^{-2}$   &  0.1&  0.2&  0.6&  1.2\\
 \hline
 2 & $N_\mathrm{d}$& 0&  0&  0&  0&  0&  1\\[-0.25cm]
(740t) &$N_{\mathrm{bk}}$&  $1\cdot 10^{-2}$ & $5\cdot 10^{-2}$    &  0.1&  0.2&  0.5&  1.0\\
 \hline
 3 & $N_\mathrm{d}$& 0&  0&  0&  0&  0&  2\\[-0.25cm]
(728t) &$N_{\mathrm{bk}}$&  $1\cdot 10^{-2}$& $5\cdot 10^{-2}$  &  0.1&  0.2&  0.5&  1.0\\
 \hline
 4 & $N_\mathrm{d}$& 0&  0&  0&  0&  0&  1\\[-0.25cm]
(726t) &$N_{\mathrm{bk}}$&  $1\cdot 10^{-2}$& $6\cdot 10^{-2}$  &  0.1&  0.2&  0.6&  1.1\\
 \hline
 5 & $N_\mathrm{d}$& 0&  0&  0&  0&  1&  2\\[-0.25cm]
(666t) &$N_{\mathrm{bk}}$&  $1\cdot 10^{-2}$&  $5\cdot 10^{-2}$  &  0.1&  0.2&  0.5& 0.9\\
 \hline
 6 & $N_\mathrm{d}$& 0&  1&  1&  1&  2&  2\\[-0.25cm]
(786t) &$N_{\mathrm{bk}}$&  $1\cdot 10^{-2}$&  $5\cdot 10^{-2}$  &  0.1&  0.2&  0.5&  1.0\\
 \hline
 7 & $N_\mathrm{d}$& 0&  0&  0&  0&  0&  0\\[-0.25cm]
(364t) &$N_{\mathrm{bk}}$&  $4\cdot 10^{-3}$&  $2\cdot 10^{-2}$&  $4\cdot 10^{-2}$&  0.1&  0.2&  0.4\\
 \hline
 8 & $N_\mathrm{d}$& 0&  0&  0&  1&  1&  1\\[-0.25cm]
(364t) &$N_{\mathrm{bk}}$&  $4\cdot 10^{-3}$&  $2\cdot 10^{-2}$&  $4\cdot 10^{-2}$&  0.1&  0.2&  0.4\\
 \hline
 \hline
 $\Sigma_1^8$ & $N_\mathrm{d}$& 0&  1&  1&  2&  5&  12\\[-0.25cm]
$\bar M$=639t &$N_{\mathrm{bk}}$&  0.07&  0.35& 0.7& 1.4 & 3.5&  7.0\\
\hline
\end{tabular}
\label{tb2}
\end{center}
\end{table}
The differences between $N_\mathrm{d}$ and $N_{\mathrm{bk}}$ are
within the statistical fluctuations, for all data classes and for
all the events. The most significant effect is observed in
correspondence of GW event n.5: 11 pulses detected against 3.4
expected, when using IBD class data and $\delta t = 100$ s. Taking
into account the number of trials ($8\times 3\times 6$), the
associated chance probability  is $P=0.03$. In order to check the
consistency of such an effect with a physical one, we complete the
coincidence analysis with the study of the time distribution of
both high and low energy signals.

\subsection{Step 4. Time distribution of pulses.}
\begin{figure}[H]
\vspace{-1.5cm}
\begin{center}
\mbox{\epsfig{file=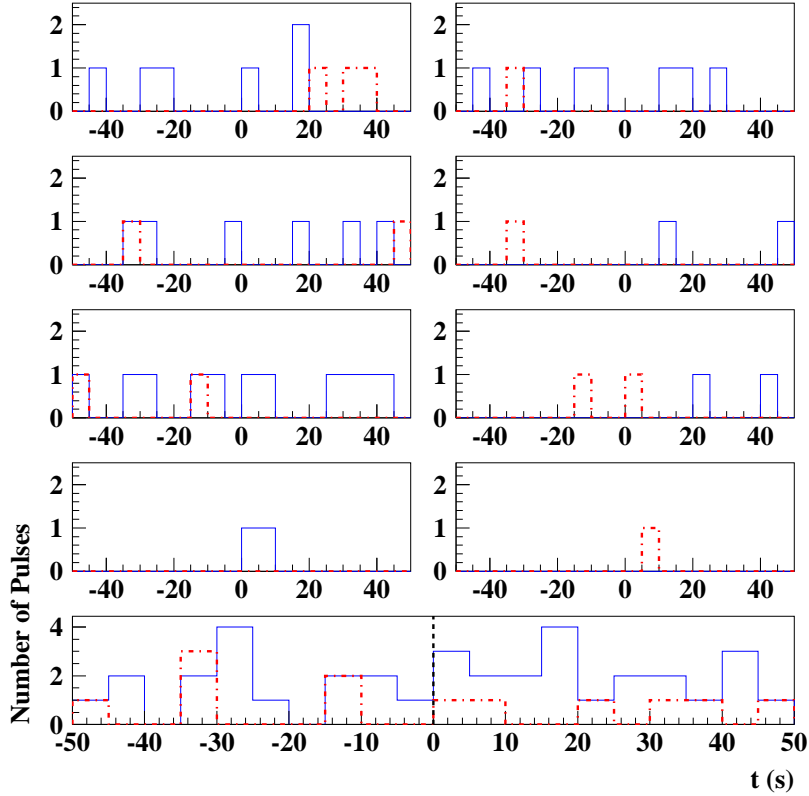,height=12cm,width=12cm}}   
\caption{Time distribution (bin=$5$s) of the detected pulses (full
line for IBD class data and dash-dotted line for NC class data)
around the corresponding GW event time ($t=0$); the bottom panel
shows the distribution for the 8 events taken together. }
\label{timepulse8}
\end{center}
\end{figure}

We have studied the time
distribution of LVD pulses around each GW event. Figure
\ref{timepulse8} shows such a distribution for each of the 8
events (and for their sum), for IBD data (full line) and NC data
(dash-dotted line), ($t=0$ corresponds to the time of the GW
event): as for all the other classes, no particular time structure
is present.
Finally, for IBD class data, we have also checked the time
distribution of secondary pulses (i.e., those possibly due to
neutron capture) with respect to the prompt ones. The measured
distribution is uniform and compatible with the one expected in
the case of pure background, where delayed and prompt signals are
uncorrelated and the distribution of the differences in time is
flat (On the contrary, if the pulses were due to $\bar\nu_e$
interactions with protons, the distribution of time delays should
show an exponential behavior, with $\tau \sim 180 \mu s$,
corresponding to the average capture time of neutrons in the LVD
counters).
\section{Calculation of Upper limits}
\subsection{Upper limits on neutrino fluence}
No evidence for any statistically relevant signal in
LVD, in the three considered reaction channels (corresponding to
different neutrino species) and over a wide range of time
durations, has been found in correspondence of any of the 8 excess
events detected in coincidence by NAUTILUS and EXPLORER.\\
In the absence of any $\nu$ signal, we calculate $90\%$ C.L.
neutrino fluence upper limits at the detector without assuming
particular energy spectra, i.e., on mono-energetic neutrinos at
different energies:
\begin{equation}
\Phi(E_{\nu})=\frac{N_{90}}{M \cdot N_t \cdot  \sigma(E_{\nu}) \cdot
\epsilon(E_{\nu}) }
\end{equation}
where: $N_{90}$ is the $90 \%$ c.l. upper limit on the number of
LVD signals per GW event, obtained following Montanet et al. 1994
in the case of Poisson processes with background. The considered
background value represents the total number of expected
background events for all the 8 GWD events; the signal value, as
well, is the total number of detected events for all the 8 GWD
events. $M$ is the detector active mass in ton (summed over the 8
events); $N_t$ is the number of targets per ton (either protons or
$^{12}C$ nuclei); $\varepsilon$ is the detection efficiency;
$\sigma(E_{\nu})$ is the appropriate cross section.
Results are shown in table \ref{tb3} for $\delta t =20s$ and $100 s$.\\
\begin{table}[H]
\vspace{-0.1cm} \caption{Fluence upper limits ($90\%$ C.L.) for
neutrinos of different energies, obtained from IBD and NC classes
of events.}
\begin{center}
\vspace{-0.2cm}
\begin{tabular}{| c || c  c || c  c |}
\hline $E_{\nu}$ & $ \delta t =20$ $s$ & & $\delta t = 100$ $s$ &\\
\hline \hline & $\Phi_{\bar\nu_e}$ $[cm^{-2}]$ & $\Phi_{\nu_i}$
$[cm^{-2}]$ & $ \Phi_{\bar\nu_e}$ $[cm^{-2}]$ & $\Phi_{\nu_i}$
$[cm^{-2}]$ \\
\hline
 10 MeV & $4.64\cdot 10^{9}$ & -- & $9.85\cdot 10^{9}$ & --\\[-0.10cm]
 \hline
 15 MeV & $1.94\cdot 10^{9}$ & -- & $4.13\cdot 10^{9}$ & --\\[-0.10cm]
 \hline
 20 MeV & $1.08\cdot 10^{9}$ & $1.29\cdot 10^{11}$ & $2.30\cdot 10^{9}$ & $3.32\cdot 10^{11}$\\[-0.1cm]
 \hline
 30 MeV & $4.97\cdot 10^{8}$ & $1.51\cdot 10^{10}$ & $1.05\cdot 10^{9}$ & $3.90\cdot 10^{10}$\\[-0.1cm]
 \hline
 40 MeV & $2.94\cdot 10^{8}$ & $6.06\cdot 10^{9}$ & $6.25\cdot 10^{8}$ & $1.56\cdot 10^{10}$\\[-0.1cm]
 \hline
 50 MeV & $2.00\cdot 10^{8}$ & $3.54\cdot 10^{9}$ & $4.25\cdot 10^{8}$ & $9.10\cdot 10^{9}$\\[-0.1cm]
 \hline
 60 MeV & $1.48\cdot 10^{8}$ & $2.50\cdot 10^{9}$ & $3.15\cdot 10^{8}$ & $6.44\cdot 10^{9}$\\[-0.1cm]
 \hline
 70 MeV & $1.16\cdot 10^{8}$ & $2.01\cdot 10^{9}$ & $2.47\cdot 10^{8}$ & $5.18\cdot 10^{9}$\\[-0.1cm]
 \hline
 80 MeV & $9.52\cdot 10^{7}$ & $1.74\cdot 10^{9}$ & $2.02\cdot 10^{8}$ & $4.48\cdot 10^{9}$\\[-0.1cm]
 \hline
 90 MeV & $7.85\cdot 10^{7}$& $1.60\cdot 10^{9}$ & $1.67\cdot 10^{8}$ & $4.12\cdot 10^{9}$\\[-0.1cm]
 \hline
 100 MeV & $6.89\cdot 10^{7}$& $1.53\cdot 10^{9}$ & $1.46\cdot 10^{8}$ & $3.93\cdot 10^{9}$\\[-0.1cm]
 \hline
 \hline
\end{tabular}
\label{tb3}
\end{center}
\end{table}
This model independent fluence can be used to test a specific
model (characterized by $\Phi_{test}(E_{\nu})$ ) by performing the
convolution:
\begin{equation}
x = \int^{100 MeV}_{10 MeV}
\frac{\Phi(E_{\nu})}{\Phi_{test}(E_{\nu})} dE_{\nu}
\end{equation}
and if the quantity x is less (greater) than 1.0, the model
predicts more (fewer) events than the event limit $N_{90}$, and is
therefore excluded (not excluded) at the $90 \%$ confidence level.
\subsection{A model dependent interpretation}

We can discuss the result of the search within two possible
simplified scenarios for neutrino production, namely (i) thermal
emission (which we will call ``cooling'')
and (ii) non thermal emission (which we will address as ``accretion'').\\
We assume that the 8 events are due to a unique kind of source and that the distance to the hypothetical sources is
$d=10$ kpc, since the 8 GWD events are consistent with a galactic origin.\\
Concerning neutrino oscillations (Dighe and Smirnov, 2000;
Takahashy et al., 2001; Aglietta et al., 2003), due to the unknown
$\nu$ oscillation parameters, i.e., $|U_{e3}|^2$ and the mass hierarchy, we
consider four different cases:\\
1 - normal mass hierarchy with adiabatic transition at the high density (H)
resonance ($NH_{ad}$);\\
2 - normal mass hierarchy with non-adiabatic transition at the H
resonance ($NH_{non-ad}$);\\
3 - inverted mass hierarchy with adiabatic transition at the H
resonance ($IH_{ad}$);\\
4 - inverted mass hierarchy with non-adiabatic transition at the H
resonance ($IH_{non-ad}$).\\
More details on the effect of neutrino oscillations in the supernova neutrino signal in LVD can be found in (LVD coll., 2004).

We refer to Appendix A for details on the calculation of the number of events in each detection channel.
\subsubsection{Cooling} In the simplified ``cooling'' process we
are considering (for example in the case of a newly formed neutron
star), neutrinos of every flavor are emitted by electron-positron
annihilation ($e^- e^+,\nu_i \bar\nu_i$, $i=e,\mu,\tau$) with
thermal spectra, that is, we are assuming zero pinching. We assume
exact equipartition of the total emitted energy $E_{B}$ among all
neutrino flavors ($E_{\nu_i}=f_{\nu_i} E_B$ with $f_{\nu_i}=1/6$)
and a hierarchy of the mean temperatures of the different flavors
$T_{\nu_x}>T_{\bar\nu_e}=T_{\nu_e}$ ($x=\mu,\tau$). The
characteristics of this emission process and the numerical values
used are summarized in tab. \ref{tb4} and in the Appendices.
\begin{table}[H]
\vspace{-0.5cm} \caption{Characteristics of the considered
processes of $\nu$ emission and numerical values used in the
limits calculation.}
\begin{center}
\begin{tabular}{c c c }
\hline Process & ``cooling'' & ``accretion''
\\ \hline
Emitted flavors & $\nu_i \bar\nu_i $ & $\bar\nu_e,\nu_e $ \\
Energy spectrum & thermal & non thermal\\
$T_{\bar \nu_e}$ & $1\div10$ MeV & $1\div10$ MeV \\
$k = T_{\nu_x}/T_{\bar \nu_e}$ & $1.3\div1.5$ & - \\
$f_{\nu_e} \equiv f_{\bar\nu_e}$ & $1/6$ & $1/2$ \\
$f_{\nu_x}$  & $1/6$ & $0$ \\
\hline
\end{tabular}
\label{tb4}
\end{center}
\end{table}
\begin{figure}[H]
\vspace{-0.5cm}
\begin{center}
\mbox{\epsfig{file=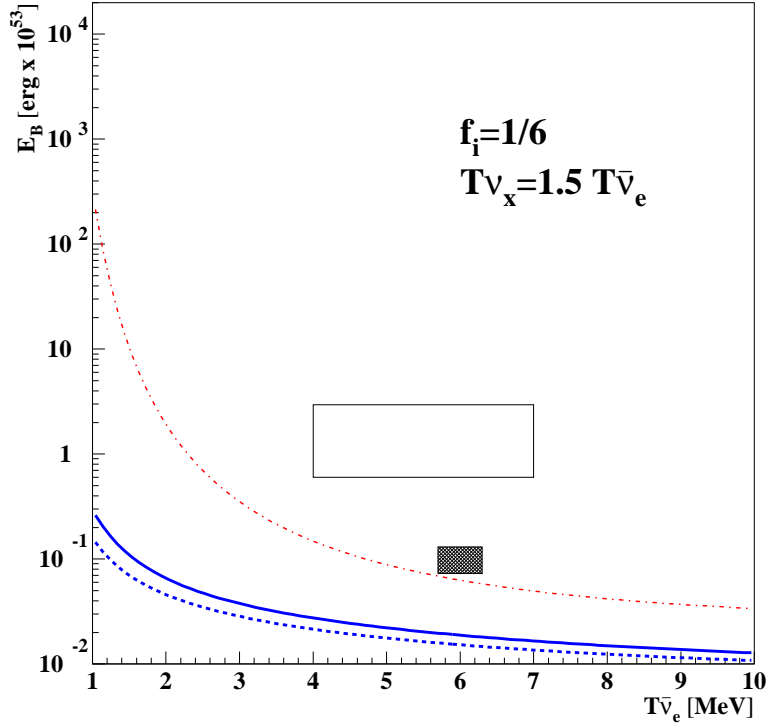,height=11cm,width=11cm}}  
\end{center}
\vspace{-0.8cm} \caption{$90 \%$ c.l. limits to the total energy
emitted in neutrinos, in the cooling process, during 20 s,
assuming pure Fermi-Dirac spectrum. The dashed line represents the
$IH_{ad}$ case, while the other three cases are represented by the
solid curve. The limit obtained through NC events is represented
by the dash-dotted line.} \label{c1}
\end{figure}
The limits obtained using IBD events - which are the most
stringent (see the appendices for the calculation) - in terms of
total emitted energy $E_B$, are shown in fig. \ref{c1} as a
function of $T_{\bar \nu_e}$, in the case of exact energy
equipartition among all flavors, with $T_{\nu_x}=1.5 T_{\bar
\nu_e}$ and $\delta t = 20 s$. \footnote{ Using the values:
$f_{\nu_e} = 1/5$ and $T_{\nu_x}=1.3 T_{\bar \nu_e}$, the limits
weakens of a factor $<2$.} The dashed line represents the
$IH_{ad}$ case, while the other three cases are not
distinguishable among them and are represented by the solid curve.
For the sake of completeness, we shown in the same figure the
limit obtained through NC events (dash-dotted line)
which is independent on $\nu$ flavor.\\
As an example for the case of cooling, we show in the same figure
the total energy expected to be emitted in two of the most
probable sources: a new-born neutron star (Keil, Raffelt and
Janka, 2002, and references therein) empty box, and colliding
neutron stars, full box (Ruffert and Janka, 1998).
\subsubsection{Accretion}
\begin{figure}[H]
\vspace{-1.5cm}
\begin{center}
\mbox{\epsfig{file=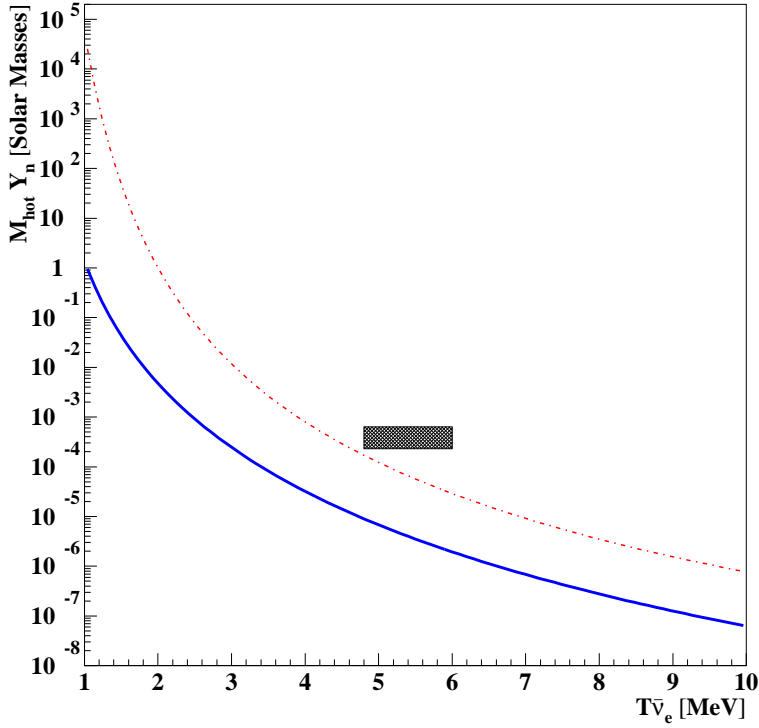,height=11cm,width=11cm}}  
\end{center}
\vspace{-0.8cm} \caption{$90 \%$ c.l. limits to the total
accretion mass, multiplied by the factor $Y_n$, as a function of
$T_{\bar \nu_e}$. The solid line represents the limits obtained
using IBD interactions in the case of $NH$ and $IH_{non-ad}$, the
dash-dotted one refer to limit derived through NC events.}
\label{c2}
\end{figure}

During matter accretion in a neutron-rich ambient, another process
appear to generate $\nu$ emission (see e.g. T.J.Loredo and
D.Q.Lamb, 2002): $e^{\pm}$ pairs, created in the accreting matter,
are captured by protons and neutrons and the resulting $\nu$
fluxes have the following characteristics: (i) only $\nu_e$ and
$\bar\nu_e$ are emitted; (ii) the thermal energy spectrum is
multiplied for a $E_{\nu}^2$ term accounting for the capture cross
section energy dependence. The characteristics of this emission
process and the numerical values are shown in tab. \ref{tb4}. The
obtained limits (see the appendices for the calculation) in terms
of accretion mass multiplied by the neutron fraction of accreting
matter, $Y_n$, are shown in Figure \ref{c2}, as a function of
$T_{\bar \nu_e}$. The solid line represents the limits obtained
using IBD interactions in the case of $NH$ and $IH_{non-ad}$.
Since for $IH_{ad}$ only a minimum part ($|U_{e3}|^2$) of the
original $\bar\nu_e$ flux will interact as $\bar\nu_e$, we do not
use the IBD events. It still remains valid the one obtained with
NC events, represented by the dash-dotted line.

As an example, we show in the same figure (full box) the amount of
accreted matter (multiplied by $Y_n=0.5$) expected for coalescing
neutron stars (Ruffert and Janka, 2001).
\section{Conclusions}
We have conducted a search for low-energy antineutrino (neutrino)
bursts with the LVD detector in coincidence with the 8 event
excess found by the gravitational waves detectors
EXPLORER and NAUTILUS during the year 2001.\\
We have found no evidence for any statistically relevant signal in
LVD, in three different reaction channels (inverse beta decay,
charged current and neutral current interactions with $^{12}C$)
corresponding to different neutrino species, over a wide range of
time durations, for any of the 8 events. Consequently, we have
derived $90 \%$ fluence upper limits to antineutrino and neutrino
emission from an average GW event, at different energies in the
range of sensitivity of the LVD detector.\\
We have then related the result of the search with two possible
simplified models for neutrino emission, i.e., ``cooling'' and
``accretion'', deriving limits, on the one side, to the total
energy emitted in neutrinos at the source, and, on the other, to
the amount of accreting mass. Assuming a source distance $d=10$
kpc, possible candidates as new-born and colliding neutron stars
have been excluded by this analysis.
This makes even more challenging and interesting the search for a
likely astrophysical source for the reported GWD events.\\[1cm]
{\it Acknowledgments}. The authors are grateful to the director
and the staff of the National Gran Sasso Laboratory for their
constant and valuable support. The authors thank precious
comments by Francesco Vissani.\\[1cm]

{\bf Appendix A. Limits calculation: inverse $\beta$ decay
interactions.}\\[0.2cm]
The number of $\bar{\nu}_e$ interactions due to inverse beta decay
in a detector is given by:
\begin{center}
\begin{equation}
N_{ev}^{IBD}=M \cdot N_{p} \cdot
\int_{Q}^{\infty}
\frac{dN_{\bar{\nu}_e}}{dE_{\bar{\nu}_e}} \cdot
\sigma(E_{\bar{\nu}_e}) \cdot \varepsilon_n \cdot
\varepsilon(E_d,E_{th}) dE_{\bar{\nu}_e} \label{Nevcoolp}
\end{equation}
\end{center}
where: $M$ is the detector active mass in ton;
$N_p=9.36\cdot10^{28}$ is the number of free protons per ton;
$\varepsilon_n$ is the neutron detection efficiency;
$\varepsilon(E_d,E_{th})$ is the $e^+$ detection efficiency;
$E_d=E_{\bar{\nu}_e}-Q+2m_ec^2$, with $Q=M_n+m_e-M_p=1.8$ MeV is
the positron detectable energy; $E_{th}$ is the detector energy
threshold; $\sigma(E_{\bar{\nu}_e})$ is the cross section (Vogel
and Beacom, 1999; Strumia and Vissani, 2003);
$\frac{dN_{\bar\nu_e}}{dE_{\bar\nu_e}}$ is the antineutrino energy spectrum.\\
In the case of cooling process it is:
\begin{center}
\begin{equation}
\frac{dN_{\bar\nu_e}}{dE_{\bar\nu_e}}= \frac{E_{B}}{4\pi d^2}
\cdot \frac{120}{7\pi^4} \cdot F_{\bar\nu_e}
\end{equation}
\end{center}
where $d=10 Kpc$ is the assumed source distance;\\ $E_{B} =
\sum_{i}f_{\nu_i}E_{B}$ is the total energy emitted in neutrinos;\\
$F_{\nu}$ is the term accounting for different $\nu$ oscillation
scenarios\footnote{NC data are not affected by oscillations.
However, the limits from IBD data stay almost the same even
assuming that MSW oscillations are completely absent. In fact, on
accounting for vacuum oscillations we get
$P_{\bar{e}\bar{e}}=1-sin^2 2\theta_{12}/2 \sim 0.6$ in all
scenarios; this is practically the same value implied by MSW
oscillations in the scenarios $NH$ and $IH_{non\ ad.}$,
$P_{\bar{e}\bar{e}}=cos^2\theta_{12}\sim 0.7$.}:
\begin{itemize}
\item for $NH_{ad}$; $NH_{non-ad}$; $IH_{non-ad}$:
\begin{center}
\begin{equation}
F_{\bar{\nu}_e}=
\frac{f_{\nu_e}}{T_{\bar{\nu}_e}^4}|U_{e1}|^2\frac{E^2}{1+e^{E/
T_{\bar{\nu}_e}}} +
\frac{f_{\nu_x}}{T_{\bar{\nu}_x}^4}|U_{e2}|^2\frac{E^2}{1+e^{E/
T_{\bar{\nu}_x}}} \label{osc1ae}
\end{equation}
\end{center}
\item for $IH_{ad}$:
\begin{center}
\begin{equation}
F_{\bar{\nu}_e}=
\frac{f_{\nu_e}}{T_{\bar{\nu}_e}^4}|U_{e3}|^2\frac{E^2}{1+e^{E/
T_{\bar{\nu}_e}}} +
\frac{f_{\nu_x}}{T_{\bar{\nu}_x}^4}(1-|U_{e3}|^2)\frac{E^2}{1+e^{E/
T_{\bar{\nu}_x}}} \label{osc2ae}
\end{equation}
\end{center}
\end{itemize}
with: $|U_{e1}|^2 \approx cos^2\theta_{12}=0.67$, $|U_{e2}|^2
\approx sin^2\theta_{12}=0.33$, $|U_{e3}|^2 \geq 10^{-4}$ for the
adiabatic case and $|Ue3|^2 \leq 10^{-6}$ for the non adiabatic one (M. Apollonio et al., 1999) (J.Bahcall and C.Pena-Garay, 2003).\\

In the case of mass accretion process it is (Loredo and Lamb,
2002):
\begin{center}
\begin{equation}
\frac{dN_{\bar\nu_e}}{dE_{\bar\nu_e}}= \frac{1}{4\pi d^2} \cdot
A_a \cdot Y_n \cdot M_{hot} \cdot F'_{\bar\nu_e}
\end{equation}
\end{center}
where:\\
$M_{hot}$ is the mass of hot emitting material;\\
$Y_n$ is the neutron fraction;\\
$A_a=\frac{1+3g^2_A}{8}\frac{\sigma_{0}c}{m_n(m_ec^2)^2}\frac{8\pi}{(hc)^3}$,
with $g_A=1.254$, $\sigma_{0}=1.7 \cdot 10^{44} cm^2$;\\
and with respect to $F'_{\nu}$:
\begin{itemize}
\item for $NH_{ad}$; $NH_{non-ad}$; $IH_{non-ad}$:
\begin{equation}
F'_{\bar{\nu}_e}= |U_{e1}|^2\frac{E^4}{1+e^{E/ T_{\bar{\nu}_e}}}
\label{osc3ae}
\end{equation}
\item for $IH_{ad}$:
\begin{equation}
F'_{\bar{\nu}_e}=|U_{e3}|^2\frac{E^4}{1+e^{E/ T_{\bar{\nu}_e}}}
\simeq 0 \label{osc4ae}
\end{equation}
\end{itemize}
\vspace{0.2cm}

{\bf Appendix B. Limits calculation: neutral current interactions}\\[0.2cm]
The number of interactions in the detector due to the neutral
current is given by:
\begin{center}
\begin{equation}
N_{ev}^{NC}=M \cdot N_{C} \cdot \varepsilon_C \cdot
\int_{15.1\
{\rm{MeV}}}^{\infty}[\frac{dN_{\bar{\nu}_i}}{dE_{\bar{\nu}_i}}
\sigma(E_{\bar{\nu}_i}) + \frac{dN_{\nu_i}}{dE_{\nu_i}}
\sigma(E_{\nu_i})] dE \label{Nevcoolcn}
\end{equation}
\end{center}
where: $N_{C}=4.24\cdot10^{28}$ is the number of $^{12}C$ nuclei
per ton; $\varepsilon_C$ is the detector efficiency for $15.1$ MeV
gamma; $\sigma (E_{\nu})$ is the cross section (M. Fukugita et
al., 1988).
\\
The neutrino energy spectrum in the case of cooling process is:
\begin{center}
\begin{equation}
\frac{dN_{\bar\nu_i}}{dE_{\bar\nu_i}}=\frac{dN_{\nu_i}}{dE_{\nu_i}}=
\frac{E_{B}}{4\pi d^2} \cdot \frac{120}{7\pi^4} \cdot F_{i}
\end{equation}
\end{center}
with
$F_i=   \frac{f_{\nu_e}}{T_{{\nu}_e}^4}\frac{E^2}{1+e^{E/
T_{{\nu}_e}}}
+ 2 \cdot \frac{f_{\nu_x}}{T_{{\nu}_x}^4}\frac{E^2}{1+e^{E/ T_{{\nu}_x}}}$\\
while, for the mass accretion case, we considered all the events
as $\bar\nu_e$ and we used:
\begin{center}
\begin{equation}
\frac{dN_{\bar\nu_i}}{dE_{\bar\nu_i}}=\frac{dN_{\bar\nu_e}}{dE_{\bar\nu_e}}=
\frac{1}{4\pi d^2} \cdot A_a \cdot Yn \cdot M_{hot} \cdot
\frac{E^4}{1+e^{E/ T_{\bar \nu_e}}}
\end{equation}
\end{center}

\end{document}